# WIDE BAND SUB-BAND SPEECH CODING USING NON-LINEAR PREDICTION [1]

*Marcos Faundez-Zanuy*

Escola Universitària Politècnica de Mataró
Universitat Politècnica de Catalunya (UPC)
Avda. Puig i Cadafalch 101-111, E-08303 MATARO (BARCELONA)
e-mail: faundez@eupmt.es  http://www.eupmt.es/veu

## ABSTRACT

In this paper we compare a wide band sub-band speech coder using ADPCM schemes with linear prediction against the same scheme with nonlinear prediction based on multi-layer perceptrons. Exhaustive results are presented in each band, and the full signal. Our proposed scheme with non-linear neural net prediction outperforms the linear scheme up to 2 dB in SEGSNR. In addition, we propose a simple method based on a non-linearity in order to obtain a synthetic wide band signal from a narrow band signal.

## 1. INTRODUCTION

Nowadays the increase in sampling rate and the number of quantization bits is present in a lot of applications. The most important examples are the Home cinema standards (Dolby Digital, THX, DTS) and the new systems for audio recording (DVD-Audio and Super Audio Compact Disc) that outperform the classical Compact Disc. Of course, the bandwidth of a speech signal is smaller than the music material, but the increase of applications that use full bandwidth (and conversions schemes from telephonic bandwidth to full bandwidth) is progressing. While compression standards for the Plain Old Telephonic Service are well established, a lot of research must be done on wide-band speech and audio signals. Non-linear speech processing is a promising approach [1] that contributes to high efficient encoders.

## 2. SUB-BAND CODER

In this paper, we evaluate a Sub-band coder with two bands: Low (L) with frequencies comprised between [0, 4000] Hz and High (H) [4000, 8000] Hz. Figure 1 shows the encoder scheme. The full band signal (sampled at 16 kHz, thus the bandwidth equals 8 kHz) is split into two bands, L and H, with half rate of the original signal (F or full). For this purpose a Quadrature Mirror Filter (QMF) is used. An ADPCM en-coder is independently applied to each band. Another version of the Sub-band ADPCM encoder (with LPC prediction) is used in the Digital dts Surround system, named Coherent Acoustics Coding System [2].

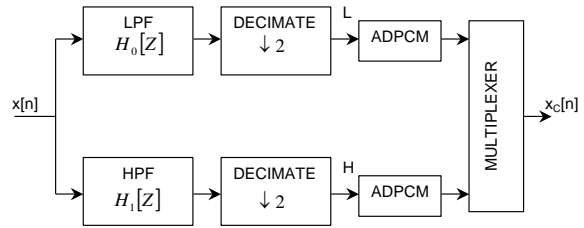

Figure1. Sub-band encoder

Figure 2 shows the decoder. Basically, it consists on the inverse blocks of the en-coder.

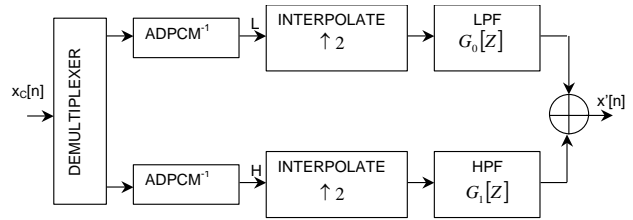

Figure 2. Sub-band decoder

### 2.1 QMF

Low and High pass filters are implemented using Quadrature Mirror Filters (QMF). QMF filters must satisfy the equation (1):

$$H_0[\omega - \pi]G_0[\omega] + H_1[\omega - \pi]G_1[\omega] = 0 \qquad (1)$$

It can be shown [3] that this can be accomplished with the following conditions (2) and (3):

$$\begin{cases} h_0[n] = h[n] \\ h_1[n] = (-1)^n h[n] \end{cases}, \qquad (2)$$

---
[1] This work has been supported by the CICYT TIC2000-1669-C04-02

$$\begin{cases} g_0[n] = 2h[n] \\ g_1[n] = -2h_1[n] = -2(-1)^n h[n] \end{cases} \quad (3)$$

We have used a 32 order FIR filter. Figure 3 shows the frequency response of the filters $H_0[\omega]$ and $H_1[\omega]$.

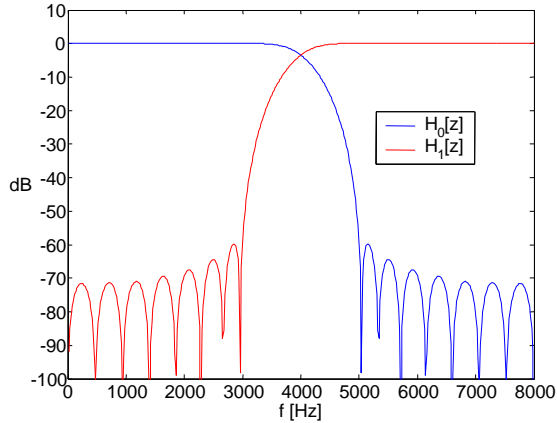

Figure 3. Frequency response of the filters

## 2.2 ADPCM

The classical ADPCM scheme has been used with linear (LPC-10, LPC-25) and nonlinear prediction. Figure 4 shows the scheme of the ADPCM scheme with non-linear prediction.

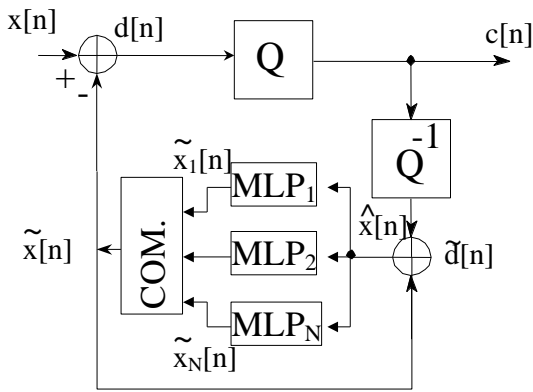

Figure 4. ADPCM scheme with non-linear prediction

Our nonlinear predictor consists on a Multi Layer Perceptron (MLP) with 10 in-puts, 2 neurons in the hidden layer, and 1 output. The selected training algorithm is the Levenberg-Marquardt, that computes the approximate Hessian matrix, because it is faster and achieves better results than the classical back-propagation algorithm. We also apply a multi-start algorithm with five random initializations for each neural net. This is represented on figure 3 with the block COM.

In [4] we studied several training schemes, and we concluded that the most suitable is the combination between Bayesian regularization and a committee of neural nets (each neural net is the result of training one random initialization).

We adapt the coefficients of the predictor on a frame (200 samples) basis, using backward adaptation: This means that the coefficients are computed over the previous frame. Thus, it is not needed to transmit the coefficients of the predictor, because the receiver has already decoded the previous frame and can obtain the same set of coefficients.

We have used adaptive scalar quantizer based on multipliers [5]. The number of quantization bits is variable between Nq=2 and Nq=5, that correspond to 32kbps and 80kbps (the sampling rate of the speech signal is 16kHz).

## 3. EXPERIMENTAL RESULTS

We have evaluated the Segmental signal to noise ratio (SEGSNR) and the Prediction Gain (Gp) defined as:

$$SNRSEG = \frac{1}{K} \sum_{j=1}^{K} SNR_j$$

, where $SNR_j$ is the signal to noise ratio (dB) of frame $j$, that is

$$SNR[dB] = 10\log_{10}\left( \frac{E\{x^2[n]\}}{E\{e^2[n]\}} \right)$$

, $K$ is the number of frames of the encoded file, and $e[n] = x[n] - \hat{x}[n]$ is the error of the decoded signal. We define analogously the prediction gain, using $d[n]$ instead of $e[n]$. Thus, $Gp$ takes into account just the predictor effect.

We have used a database of 1 sentence extracted from the Gaudi-Ahumada database [6]. Table 1 shows the results for linear prediction of order 10, and table 2 for linear prediction of order 25. This results can be compared against the same scheme with neural net prediction (prediction order=10, number of coefficients=25). Table 3 shows the results for the Multi Layer Perceptron predictor.

| | B | Nq=2 | | Nq=3 | | Nq=4 | | Nq=5 | |
|---|---|---|---|---|---|---|---|---|---|
| | | m | σ | m | σ | m | σ | m | σ |
| $G_p$ | L | 9.40 | 6.4 | 11.32 | 5.4 | 11.61 | 5.6 | 11.65 | 5.6 |
| | H | -0.15 | 1.0 | -0.02 | 1.1 | -0.02 | 1.1 | -0.02 | 1.1 |
| SEG | L | 17.09 | 6.6 | 20.62 | 8.2 | 25.15 | 8.0 | 28.73 | 8.5 |
| | H | 7.30 | 1.1 | 12.63 | 1.3 | 17.00 | 1.7 | 21.45 | 2.1 |
| | F | 16.84 | 6.6 | 22.60 | 6.8 | 26.77 | 6.8 | 30.91 | 6.9 |

Table 1. Mean (m) and standard deviation (σ) of the Prediction Gain (Gp) and SEGSNR (SEG) for several ADPCM schemes with LPC-10, backward adaptation. B= Band, L= Low, H= High, F= Full.

It is important to take into account that the SEGSNR and Gp measures of the L and H signals have been obtained using only the L or H band, considering as reference ("original" signal) the respective output of their filter, before ADPCM encoding.

|   | B | Nq=2 | | Nq=3 | | Nq=4 | | Nq=5 | |
|---|---|------|---|------|---|------|---|------|---|
|   |   | m | σ | m | σ | m | σ | m | σ |
| $G_p$ | L | 8.80 | 6.3 | 10.93 | 5.5 | 11.05 | 5.6 | 11.23 | 5.7 |
|   | H | -0.52 | 1.1 | -0.42 | 1.2 | -0.43 | 1.2 | -0.44 | 1.2 |
| SEG | L | 16.45 | 6.4 | 20.44 | 8.1 | 23.95 | 8.0 | 28.42 | 8.6 |
|   | H | 6.93 | 1.2 | 12.26 | 1.5 | 16.59 | 2.0 | 20.90 | 2.5 |
|   | F | 16.22 | 6.4 | 22.21 | 6.8 | 26.40 | 7.0 | 30.71 | 7.1 |

Table 2. Mean (m) and standard deviation (σ) of the Prediction Gain (Gp) and SEGSNR for several ADPCM schemes with LPC-25, backward adaptation.

|   | B | Nq=2 | | Nq=3 | | Nq=4 | | Nq=5 | |
|---|---|------|---|------|---|------|---|------|---|
|   |   | m | σ | m | σ | m | σ | m | σ |
| $G_p$ | L | 8.84 | 6.4 | 10.32 | 7.0 | 10.39 | 7.2 | 10.42 | 7.1 |
|   | H | 0.18 | 0.9 | 0.19 | 1.0 | 0.21 | 1.0 | 0.21 | 1.0 |
| SEG | L | 16.57 | 6.5 | 23.51 | 7.1 | 28.19 | 7.4 | 33.05 | 7.5 |
|   | H | 7.61 | 1.0 | 13.03 | 1.4 | 17.32 | 1.8 | 21.77 | 2.5 |
|   | F | 16.55 | 6.2 | 23.08 | 7.0 | 27.97 | 7.4 | 32.72 | 7.6 |

Table 3. Mean (m) and standard deviation (σ) of the Prediction Gain (Gp) and SEGSNR for several ADPCM schemes with nnet prediction, backward adaptation.

The number of bits in each band can de setup independently, so there are several possibilities. Table 4 shows the different alternatives using quantizers ranging from 2 to 5 bits per sample, and the equivalent bit rate of the full signal.

| L \ H | Nq=2 | Nq=3 | Nq=4 | Nq=5 |
|-------|------|------|------|------|
| Nq=2 | 2 | 2.5 | 3 | 3.5 |
| Nq=3 | 2.5 | 3 | 3.5 | 4 |
| Nq=4 | 3 | 3.5 | 4 | 4.5 |
| Nq=5 | 3.5 | 4 | 4.5 | 5 |

Table 4. Bit rate as function of the low (L) and high (H) band bit rates

Tables 5 and 6 show the experimental results for lineal prediction (prediction orders 10 and 25 respectively), and table 7 for nonlinear prediction. Best results for each bit rate are under-lined.

From these tables it can be deduced that for a given bit rate, it is preferred to assign more bits to the L band. For instance, for L and H, 4 and 2 bits respectively, rather than 3+3 or 2+4. This is consistent with the fact that most part of the energy and information of a speech signal are comprised between [0, 4 kHz] range, that is the L band. This result also holds for the non-linear prediction scheme.

| | LPC-10 | | | | | | | |
|---|---|---|---|---|---|---|---|---|
| H \ L | Nq=2 | | Nq=3 | | Nq=4 | | Nq=5 | |
|   | m | σ | m | σ | m | σ | m | σ |
| Nq=2 | 16.8 | 6.5 | 17.5 | 6 | 17.7 | 5.8 | 17.8 | 5.7 |
| Nq=3 | 21.2 | 8 | 22.6 | 6.8 | 23.2 | 6.3 | 23.4 | 6.1 |
| Nq=4 | 23.8 | 9.1 | 25.8 | 7.6 | 26.8 | 6.8 | 27.3 | 6.3 |
| Nq=5 | 25.7 | 10 | 28.3 | 8.7 | 29.8 | 7.7 | 30.9 | 6.9 |

Table 5. SEGSNR of F as function of the low (L) and high (H) band bit rates and LPC-10

| | LPC-25 | | | | | | | |
|---|---|---|---|---|---|---|---|---|
| H \ L | Nq=2 | | Nq=3 | | Nq=4 | | Nq=5 | |
|   | m | σ | m | σ | m | σ | m | σ |
| Nq=2 | 16.2 | 6.5 | 16.8 | 5.9 | 17.1 | 5.8 | 17.1 | 5.7 |
| Nq=3 | 20.9 | 7.9 | 22.2 | 6.8 | 22.7 | 6.3 | 23 | 6.1 |
| Nq=4 | 23.5 | 9.1 | 25.4 | 7.7 | 26.4 | 6.9 | 26.9 | 6.5 |
| Nq=5 | 25.6 | 10 | 28.2 | 8.8 | 29.7 | 7.8 | 30.7 | 7.1 |

Table 6. SEGSNR of F as function of the low (L) and high (H) band bit rates and LPC-25

| H \ L | Nq=2 | | Nq=3 | | Nq=4 | | Nq=5 | |
|---|---|---|---|---|---|---|---|---|
|   | m | σ | m | σ | m | σ | m | σ |
| Nq=2 | 16.55 | 6.2 | 17.10 | 5.7 | 17.26 | 5.5 | 17.34 | 5.5 |
| Nq=3 | 21.81 | 8.1 | 23.08 | 7.0 | 23.58 | 6.6 | 23.82 | 6.3 |
| Nq=4 | 25.03 | 9.6 | 27.02 | 8.2 | 27.97 | 7.4 | 28.52 | 6.9 |
| Nq=5 | 27.32 | 10.9 | 30.13 | 9.5 | 31.67 | 8.4 | 32.72 | 7.6 |

Table 7. SEGSNR of F as function of the low (L) and high (H) band bit rates and MLP

## 4. SYNTHETIC HIGH FREQUENCY BAND GENERATION

Another approach for wide band coding is the elimination of the high frequency portion of the spectrum [4000,8000] Hz, and to recover this band at the decoder with a bandwidth extension (or bandwidth replication) algorithm. We propose a new simple method that consists on the generation of high frequency harmonics by means of a quadratic non-linearity. It is important to take into account that the low frequency band [0, 4000] Hz remains the same, because the artificial high frequency signal is high pass filtered and added to the low frequency band. The use of a QMF filters for H0 and H1 (see figure 3) avoids phase problems.

Figure 5 shows the proposed scheme. Of course, it is possible to use more complicated schemes, such as the proposed in [7], that has been successfully evaluated in combination with speaker identification [8] and verification [8].

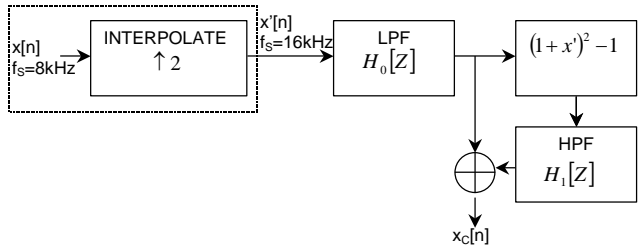

Figure 5. Synthetic High frequency band generation.

Listening tests shows that the wide-band signal obtained with the scheme of figure 5 sounds more natural than his equivalent narrow band signal, although the bit rate is the same for both signals (no high band frequency information is encoded).

Figure 6 shows an example of the spectrum for a voiced frame in several situations (narrow band, real full band, and synthetic full band).

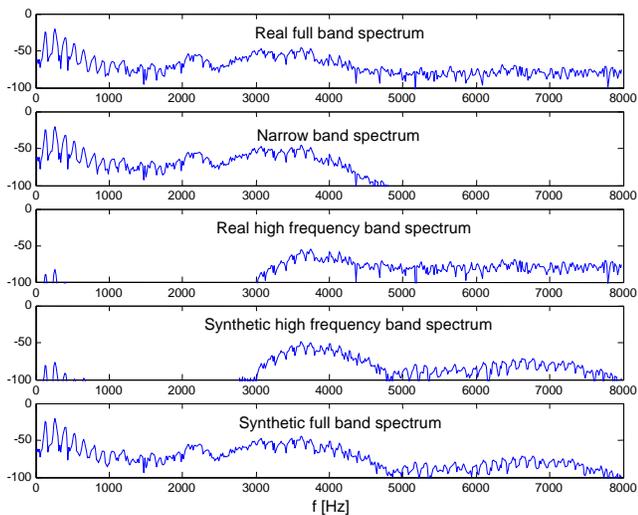

Figure 6. Example of narrow band, real full band, and synthetic full band spectrums for a voiced frame.

## 5. CONCLUSIONS

In this paper, we have presented a new wide-band speech coding scheme with non-linear prediction based on Multi Layer Perceptrons. The numerical experiments reveal an improvement in the segmental signal to noise ratio of up to 2 dB when the proposed non-linear scheme is compared against a linear LPC-10, LPC-25 scheme.

In addition, we have also proposed the generation of the High frequency band using a non-linearity, because it is able to reproduce the harmonic structure of periodic spectrums. The motivation of this second proposal has been the observation that little improvement can be obtained (measured with the segmental signal to noise ratio) with the coding of High frequency band. It seems that some harmonic structure on this band is enough in order to produce a more comfortable signal.